\documentstyle[12pt]{article}

\newcommand{\EQ}{\begin{equation}}
\newcommand{\EN}{\end{equation}}
\newcommand{\bea}{\begin{eqnarray}}
\newcommand{\ena}{\end{eqnarray}}

\renewcommand{\a}{\alpha}
\renewcommand{\b}{\beta}

\renewcommand{\d}{\delta}

\newcommand{\pa}{\partial}

\newcommand{\G}{\Gamma}

\renewcommand{\l}{\lambda}

\renewcommand{\S}{\Sigma}

\renewcommand{\o}{\omega}

\begin{document}
 \def\bq{\begin{quote}}
\def\eq{\end{quote}}
\topmargin -1.2cm
\oddsidemargin 5mm

\renewcommand{\Im}{{\rm Im}\,}
\newcommand{\NP}[1]{Nucl.\ Phys.\ {\bf #1}}
\newcommand{\PL}[1]{Phys.\ Lett.\ {\bf #1}}
\newcommand{\NC}[1]{Nuovo Cimento {\bf #1}}
\newcommand{\CMP}[1]{Comm.\ Math.\ Phys.\ {\bf #1}}
\newcommand{\PR}[1]{Phys.\ Rev.\ {\bf #1}}
\newcommand{\PRL}[1]{Phys.\ Rev.\ Lett.\ {\bf #1}}
\newcommand{\MPL}[1]{Mod.\ Phys.\ Lett.\ {\bf #1}}
\renewcommand{\thefootnote}{\fnsymbol{footnote}}

\newpage
\begin{titlepage}
\begin{flushright}
{IFUM 421/FT}
\end{flushright}
\vspace{2cm}
\begin{center}
{\bf{{\large SUPERSYMMETRIC, INTEGRABLE TODA}}}\\
\vspace{.1in}
{\bf{{\large   FIELD THEORIES: THE B(1,1) MODEL}}} \\
\vspace{1.5cm}
{ S. PENATI}~~ and~~ { D. ZANON} \\
\vspace{3mm}
{\em Dipartimento di Fisica dell' Universit\`{a} di Milano and} \\
{\em INFN, Sezione di Milano, I-20133 Milano, Italy}\\
\vspace{1.1cm}
{{\bf{ABSTRACT}}}
\end{center}
\bq
We study the two-dimensional supersymmetric Toda theory based on
the Lie superalgebra
$B(1,1) \equiv Osp(3|2)$ and construct its quantum W-currents. We also
investigate the fermionic affinization of this model: we show that despite
the non-unitary form of the Lagrangian the
$B^{(1)}(1,1)$ theory has a real particle mass spectrum which is not
renormalized at one-loop. We construct the first higher--spin conserved
current, prove its conservation to all-loop order, compute one-loop
corrections to the corresponding charge and check consistency between
charge and mass renormalization.
\eq
\vfill
\begin{flushleft}
IFUM 421/FT\\
May 1992\\
\end{flushleft}
\end{titlepage}
\renewcommand{\thefootnote}{\arabic{footnote}}
\setcounter{footnote}{0}
\newpage

Affine Toda theories
are massive systems obtained as perturbation of the corresponding conformal
Toda theories. The perturbation is chosen in such a way that these models
possess an infinite number of conserved currents \cite{curr,OT} and
therefore retain the
integrability properties they have at their fixed, conformal points. The
existence of higher--spin conserved charges implies elasticity and
factorization
of the S-matrix which can be determined exactly using unitarity and a bootstrap
principle \cite{zam}.
Therefore in these cases, by studying perturbed conformal field
theories one can find all the on-shell informations of the massive theories
which are encoded in their S-matrix. This program has been completed for
all {\em unitary} affine Toda theories based on simply-laced \cite{simply}
as well as nonsimply-laced Lie algebras \cite{nonsimply}:
the quantum integrability has been established
and the exact S-matrices have been constructed.

Unitary fermionic extensions of these
models \cite{ols}, i.e. the affine Toda theories based on the Lie
superalgebras $A^{(2)}(0,2n-1)$, $C^{(2)}(n+1)$, $B^{(1)}(0,n)$,
$A^{(4)}(0,2n)$, have been considered also,
and the construction of their exact S-matrices has been successfully carried
out \cite{dgpz,nonsimply}.
Despite the presence of fermions in the spectrum these theories are not
supersymmetric.

If one insists on supersymmetry, one is led to consider
affine Toda theories which are {\em not} manifestly unitary \cite{ols,EH}.
In the untwisted cases these are the affine Toda theories associated to the
purely fermionic admissible root systems of the $A^{(1)}(n,n)$,
$B^{(1)}(n,n)$, $D^{(1)}(n,n-1)$,
and $D^{(1)}(2,1;\a)$ superalgebras \cite{kac,sorba}.
The general understanding of such theories
is incomplete, however there is evidence to suggest
that for special values of the coupling constant they might admit a
unitary restriction and describe
the perturbation of "minimal" conformal field theories \cite{min}.
For some specific bosonic systems
these conjectures are supported by the fact that
even if the Lagrangian is not manifestly unitary the theory has a particle
and soliton spectrum which is real and stable under perturbation \cite{holl}.

In this paper we consider the supersymmetric theories: we
study as an explicit example the supersymmetric affine Toda theory based on the
Lie superalgebra $B^{(1)}(1,1)$ and address the issue of its quantum
integrability.
A more general analysis will be presented
in Ref. \cite{gpz2}.

We work in Minkowski space with light-cone coordinates
\bea
 z \equiv x^+ =\frac{1}{\sqrt{2}} (x^0 + x^1) \qquad \qquad
\bar{z}\equiv x^-= \frac{1}{\sqrt{2}}(x^0 - x^1) \nonumber \\
\partial \equiv \pa_+ =\frac{1}{\sqrt{2}} (\partial_0 + \partial_1) \qquad
\qquad \bar{\partial} \equiv \pa_-= \frac{1}{\sqrt{2}}
(\partial_0 - \partial_1) \qquad \qquad
\Box =2 \pa \bar{\pa}
\ena
Since the theory we will study is supersymmetric, it is convenient to perform
all the calculations in N=1 superspace with
 coordinates $Z=(z,\bar{z}, \theta, \bar{\theta})$ and
supercovariant spinor derivatives
\EQ
D = \partial_{\theta} + i \theta \partial  \qquad \qquad
\bar{D} = \partial_{\bar{\theta}} - i \bar{\theta} \bar{\partial}
\EN
satisfying the anticommutation relations $ \{ D, \bar{D}\}=0$ and
$D^2 = i \partial$, $\bar{D}^2 =-i \bar{\partial}$.
Bosonic and fermionic fields
are then components of superfields,
$
\Phi \equiv \phi + \frac{1}{\sqrt{2}} \theta \psi + \frac{1}{\sqrt{2}}
\bar{\theta} \bar{\psi} + \theta \bar{\theta} F
$,
$\psi$ and $\bar{\psi}$ being Majorana--Weyl fermions.

Two-dimensional Toda theories associated to a
Lie superalgebra with $b$ bosonic and $f$ fermionic simple roots are
represented
by actions of the form \cite{ols}
\EQ
S = \frac{1}{\b^2} \int d^2 z d^2 \theta \left[ D\vec{\Phi} \cdot \bar{D}
\vec{\Phi} + \theta \bar{\theta} \sum_{i=1}^{b} q_i e^{\vec{\a}_i \cdot
\vec{\Phi}} + \sum_{i=1}^{f} q_i  e^{\vec{\a}_i \cdot \vec{\Phi}} \right]
\label{1}
\EN
where $\vec{\a}_i$ are the simple roots of the superalgebra, $q_i$ are the
Ka$\check{c}$ labels, $\b$ is the coupling constant and
 $\vec{\Phi} = (\Phi_1, \cdots
\Phi_r)$ is a set of $r=b+f$ superfields.

In general the system of simple roots (or equivalently the Dynkin diagram)
associated to a Lie superalgebra is not unique. Unequivalent
sets of simple roots can be obtained from a given one by acting on it with a
generalized Weyl transformation associated to fermionic roots \cite{sorba}.
Different sets of simple roots have different content of fermionic roots.
Therefore different Toda actions can be obtained from a given Lie superalgebra.
For superalgebras which admit a set of purely fermionic simple roots
a manifestly supersymmetric Toda action can be constructed, since in this case
the term $\theta \bar{\theta}$ in eq.(3) is absent.

We consider here the supersymmetric Toda theory associated to the Lie
superalgebra $B(1,1) \equiv {\rm Osp}(3|2)$. The fermionic Dynkin diagram
of this rank--2 superalgebra is shown in Fig.$1$.
An explicit realization of the two fermionic roots can be
given in terms
of complex vectors $\vec{\a}_1=(1,-i)$, $\vec{\a}_2 =(0,i)$
with scalar products conventionally defined as $\vec{\a}_i \cdot \vec{\a}_j
= \sum_k \a_i^k \a_j^k$ (no complex conjugation).
The corresponding supersymmetric Toda action is
\EQ
S = \frac{1}{\b^2} \int d^2 z d^2 \theta \left[ D \Phi_1 \bar{D} \Phi_1 +
D \Phi_2 \bar{D} \Phi_2 + e^{\Phi_1 - i\Phi_2} + 2 e^{i \Phi_2} \right]
\label{2}
\EN
The theory is classically integrable.
The first {\em{d}} classical conserved currents can be
constructed by using the general procedure based on the Miura transformation
\cite{fl,EH},
where {\em{d}} are the dimensions of the vector representation of the algebra.
For $B(1,1)$ the $W^{(s)}$ currents are given by
\EQ
(D + D \vec{\Phi} \cdot \vec{{\l}_5})(D + D \vec{\Phi} \cdot \vec{{\l}_4})
\cdots
(D + D \vec{\Phi} \cdot \vec{{\l}_1})=
\sum_{i=0}^{5} W^{(2-i/2)}(z,\theta) D^i
\label{5}
\EN
where
$\vec{\l}_j$, $j=1,...,5$ are the weights of the vector representation,
$ \vec{\l}_1 =
-\vec{\l}_5 = \vec{\a}_1 + \vec{\a}_2$, $\vec{\l}_2 = -\vec{\l}_4 = \vec{\a}_2$
and $\vec{\l}_3 = 0$. Explicitly one obtains
\bea
(D - D\Phi_1)(D -iD\Phi_2)D(D + iD\Phi_2)(D + D\Phi_1) =
W^{(2)} + W^{(\frac{3}{2})} D + W^{(1)} D^2 +D^5 \nonumber \\
{}~~~~~~
\ena
with
\bea
W^{(1)} &=&
-i D \Phi_1 \pa \Phi_1 -i D \Phi_2 \pa \Phi_2 + D\pa \Phi_2 +2i D \pa \Phi_1
\nonumber \\
W^{(\frac{3}{2})} &=&
- i \pa^2 \Phi_2 +i D \Phi_1 D\pa \Phi_1 - (\pa \Phi_2)^2 -i D \Phi_2
D\pa \Phi_2 \nonumber \\
{}~~~~&~& +2D \Phi_1 D\pa \Phi_2  -2iD \Phi_1 D \Phi_2  \pa \Phi_2
\nonumber \\
W^{(2)} &=&
-D \pa^2\Phi_1 +D\Phi_1 \pa^2\Phi_1-i D\Phi_1 \pa^2\Phi_2 +
iD \pa \Phi_2 \pa\Phi_1
-(\pa \Phi_2)^2  D \Phi_1 \nonumber \\
{}~~~~&~&+ \pa \Phi_2 D \Phi_2 \pa \Phi_1 -i D \Phi_2
D \pa \Phi_2 D \Phi_1 \nonumber \\
&=& \frac{1}{2} D(W^{(\frac{3}{2})} +DW^{(1)})
\ena
It is straightforward to check using the equations of motion from the action
in eq.(4), that $\bar{D}$ commutes with the differential operator in
eq.(5), i.e. the currents $W^{(s)}$ are superholomorphic
\EQ
\bar{D} W^{(s)}=0
\EN
thus ensuring the classical integrability of the $B(1,1)$ system. In
particular,
 being $W^{(1)}$ proportional to the stress-energy tensor, the
action in eq.(4) describes a massless supersymmetric model which is conformally
invariant.
These properties hold at the quantum level too: the holomorphic currents
in eq.(7) maintain their form
albeit the coefficients of the various terms acquire a coupling constant
dependence as we now show.

The quantization
and renormalization of the conservation laws in eq.(8) could be studied using
for example
light-cone quantization procedures as described in  Ref. \cite{man};
however since we are mainly interested in
the affine extension of the $B(1,1)$ model, we follow here the approach
of Ref.\cite{dgz} which applies to conformal as well as
massive systems. We extend to
superspace the techniques introduced in Ref.\cite{dgz} for the
corresponding analysis
of higher--spin bosonic currents. The quantum Lagrangian is defined by
normal ordering the exponentials so that the theory is free of any ultraviolet
divergences. We use superspace propagators
\EQ
\left \langle
\Phi_i(Z,\bar{Z}) \Phi_j(0,0)\right \rangle =
- \d_{ij} \frac {\b^2}{4\pi} \bar{D} D[log(2z \bar{z})
\d^{(2)}(\theta)]
\EN
and look for potential anomalies in
\EQ
\bar{D}_Z \left \langle W^{(s)}(Z,\bar{Z}) \right \rangle \equiv \bar{D}_Z
\left \langle W^{(s)}(Z,\bar{Z})
\exp \left(\frac{i}{\b^2}
\int d^2w d^2 \theta' {\cal L}_{int} \right) \right \rangle _0
\EN
We compute local contributions to the above expression  and see if we can
cancel them by adding to the classical currents coupling constant dependent
corrections.
The Wick contractions  in eq.(10) produce a
number of $D$ and $\bar{D}$  from the various terms in the
currents and from the superspace propagators: one first reduces them using
their commutation relations and then the D-algebra is performed loop
by loop using $\bar{D} D \d^{(2)}( \theta -\theta')_{|\theta= \theta'} =1$.
As for the bosonic calculation \cite{dgz}, local contributions can arise only
by expanding the exponential in eq.(10) to first order in $ {\cal L}_{int}$;
once the D-algebra has been performed one obtains terms of the form
\EQ
\bar{D}_Z \int d^2 w {\cal A}(z,\bar{z},\theta, \bar{\theta})
\bar{D}_Z \frac {1} {(z-w)^n}  {\cal{B}}(w, \bar{w}, \theta, \bar{\theta})
\EN
where ${\cal A}$, ${\cal B}$ are products of superfields and their
$D$-derivatives. Finally a local contribution
is produced using
\EQ
\bar{D}_Z \bar{D}_Z \frac{1}{(z-w)^n}= -i \bar{\pa}_z \frac{1}{(z-w)^n}
=\frac{2\pi}{(n-1)!} \pa^{n-1}_w \d^{(2)}(z-w)
\EN
For the $W^{(1)}$ current one obtains one-loop contributions from Wick
contracting the
$D \Phi_1 \pa \Phi_1 $ and $D\Phi_2 \pa \Phi_2$ terms in the current with
the exponentials in the interaction Lagrangian. The $O(\b^2)$ corrections
are
\bea
&&\bar{D} \left \langle W^{(1)} \left(  \frac{i}{\b^2} \int d^2w d^2 \theta'
e^{\Phi_1
 - i \Phi_2} \right) \right \rangle \leadsto 0\nonumber \\
&&\bar{D} \left \langle W^{(1)} \left(  \frac{i}{\b^2} \int d^2w d^2 \theta'
e^{i \Phi_2}\right)  \right \rangle \leadsto \frac{\b^2}{8\pi}\pa \Phi_2
e^{i\Phi_2}
\ena
which can be cancelled renormalizing the terms
$D\pa \Phi_1$ and $D\pa \Phi_2$ in the classical current, thus obtaining the
quantum holomorphic stress-energy tensor
\EQ
W^{(1)}= -i D\Phi_1 \pa \Phi_1- i D\Phi_2 \pa \Phi_2  +(1- \frac{\b^2}{4 \pi})
D\pa \Phi_2 + 2i(1- \frac{\b^2}{8 \pi})D \pa \Phi_1
\EN
In the same way one can compute the quantum corrections to the
$W^{(\frac{3}{2})}$ current. In this case since the current contains
up to three factors of fields the Wick contractions lead to contributions
up to two loops, which actually vanish because of the D-algebra.
The calculation is not too difficult and one obtains
\bea
W^{(\frac{3}{2})}&=& -i (1-\frac{\b^2}{4\pi}) \pa^2 \Phi_2
+i(1-\frac{\b^2 }{2\pi}) D \Phi_1 D \pa \Phi_1 - (\pa \Phi_2)^2
-i D \Phi_2 D \pa \Phi_2 \nonumber \\
&~&+2(1-\frac{\b^2}{4\pi}) D\Phi_1 D \pa \Phi_2 -2i D \Phi_1 D \Phi_2
\pa \Phi_2
\ena
The corresponding calculation for spin $s=2$ needs not be performed
since the $W^{(2)}$ current is linearly dependent on  $W^{(1)}$ and
$W^{(\frac{3}{2})}$ (see eq.(7)) and therefore its renormalized
expression is easily obtained using eqs.(14) and (15).

We turn now to the study of the supersymmetric
massive perturbation of the
system in eq.(4). It is obtained by a fermionic
affinization of the $B(1,1)$ superalgebra \cite{kac,sorba}
which corresponds to the addition of the lowest {\it fermionic} root
$\vec{\a}_0 = -(\vec{\a}_1 + 2 \vec{\a}_2)= (-1,-i)$.
The affine $B^{(1)}(1,1)$ Toda action is given by
\EQ
S = \frac{1}{\b^2} \int d^2 x d^2 \theta \left[ D \vec{\Phi} \cdot
\bar{D} \vec{\Phi} + e^{\Phi_1 - i\Phi_2} + 2 e^{i\Phi_2} +
e^{-\Phi_1 - i\Phi_2} \right]
\label{4}
\EN
where the Ka$\check {c}$ labels have been chosen so that the one-point
functions
vanish.

Expanding the exponentials
up to third order, we obtain the classical {\em {mass}} spectrum
and the 3-point coupling of the theory :
\EQ
{\cal{L}}^{(2)} \equiv -M_1 \Phi_1^2 - M_2 \Phi_2^2 = \Phi_1^2 - 2\Phi_2^2
{}~~~~~~~~,~~~~~~~~{\cal{L}}^{(3)} =-i \Phi_1^2 \Phi_2
\EN
It follows that the bosonic masses are ($m_j^2 = 2 M_j^2$)
\EQ
m_1^2 = 2 \qquad \qquad \qquad  m_2^2 = 8
\label{mass}
\EN
Despite the manifest non-unitarity of the theory the classical
mass spectrum is real. We note that since $m_2 = 2m_1$ the two particles
are at threshold and divergent contributions, similar to the ones discussed
in Ref. \cite{gpz3}, could be produced in the on-shell
amplitudes. However one can show that wave-function
renormalization is sufficient to render the theory
free of any threshold singularities.

We want to establish now the existence of quantum
higher--spin conserved currents for the $B^{(1)}(1,1)$ model.
Since the theory contains a mass scale,
the stress-energy tensor acquires a non-vanishing trace.
It satisfies a conservation law of the form
\EQ
\bar{D} J^{(s)} + D \bar{J}^{(s)}=0
\EN
with $s=1$. The most efficient way to analyze the situation for $s>1$ is to
use the massless
perturbation techniques introduced above, with superfield propagators
as in eq.(9) and treating the whole exponentials in eq.(16) as interaction
terms \cite{dgz}.
We compute
\EQ
\bar{D}_Z \left \langle J^{(s)}(Z,\bar{Z}) \right \rangle \equiv \bar{D}_Z
\left \langle J^{(s)}(Z,\bar{Z})
\exp \left(\frac{i}{\b^2}
\int d^2w d^2 \theta' {\cal L}_{int} \right) \right \rangle _0
\EN
An anomaly would spoil the conservation law if the computation
of the r.h.s. of eq.(20) would produce {\em local}
terms which were not expressible
as $D$-derivatives of some appropriate $\bar{J}$. Since we are not interested
in the actual form of $\bar{J}$ we discard total $D$-derivatives, freely
integrating by parts on $z,\theta$. Moreover any current of the form
$J^{(s)}= D {\cal J}^{(s-\frac{1}{2})}$ is not relevant since it trivially
satisfies eq.(19). In this manner we find
that the $B^{(1)}(1,1)$ perturbed system does not have
conserved currents of spin $s= \frac{3}{2}$ or $s=2$. In particular
the $W^{(\frac{3}{2})}$ current is not conserved in the affine case
because it does not respect the symmetry
$\Phi_1 \rightarrow - \Phi_1$ of the action in eq.(16), while
$W^{(2)}$ is trivial being a total $D$-derivative.

The first nontrivial higher--spin current
appears at $s=3$, in accordance with the general statement proven for
purely bosonic affine Toda theories,
that the spins of the conserved charges should
be given by the exponents of the algebra modulo the Coxeter number
\cite{OT,cox}.
Indeed in this case one has $s=1,3$ mod $4$.

In order to study the renormalization of the spin-$3$ current
we proceed as follows: we consider
up to total $D$-derivative the most
general expression, even in $\Phi_1$ on the basis of the
$\Phi_1 \rightarrow - \Phi_1$ symmetry,
\bea
J^{(3)} &=& a D\pa \Phi_1 \pa^2 \Phi_1 +b D \pa \Phi_2 \pa^2 \Phi_2 +
c D \Phi_1 \pa \Phi_1 \pa^2 \Phi_2 +
+ d(\pa \Phi_1)^2 D \pa \Phi_2 \nonumber \\
&+& e (\pa \Phi_1 )^3 D \Phi_1 +
f ( \pa \Phi_2)^3 D \Phi_2 + g D \Phi_1 \pa \Phi_1 (\pa \Phi_2)^2
\nonumber \\
&+& h D\Phi_1 D \pa \Phi_1 D \Phi_2 \pa \Phi_2 +
k (\pa \Phi_1)^2 \pa \Phi_2 D \Phi_2
\ena
and compute all the local contributions which arise from Wick contracting with
the interaction Lagrangian. Dropping total $D$-derivatives and using various
identities valid up to integration by parts,
after a fairly amount of algebra we obtain ($\a \equiv \frac{\b^2}{2 \pi}$)
\vskip 5pt
\bea
&& \bar{D} \left \langle J^{(3)}
\left( \frac{i}{\b^2} \int d^2w d^2 \theta' e^{2i\Phi_2} \right) \right \rangle
\nonumber\\
&& \leadsto \left(
\left[ b+(1- \frac{9\a}{4}  + \frac {\a^2}{4})f \right] \pa^2
\Phi_2 \pa \Phi_2
+\left[- \frac{1}{2} c +i(1- \frac{\a}{4})g
+(\frac{1}{2} - \frac{\a}{4})h \right] D\Phi_1
D \pa^2 \Phi_1 \right. \nonumber \\
&& \left.
+\left[ic+id +(2-\frac{\a}{2})g+
(1- \frac{\a}{2})k  \right] \pa \Phi_1 \pa^2 \Phi_1
+ g D \Phi_1 \pa \Phi_1 D \pa \Phi_2 \right) e^{2i\phi_2}
\ena
\vskip 4pt
\noindent
and
\vskip 2pt
\bea
&&\bar{D} \left \langle J^{(3)} \left( \frac{i}{\b^2}
\int d^2w d^2\theta' e^{\Phi_1-i\Phi_2} \right) \right \rangle  \nonumber\\
&&~~~~\leadsto
\left( \left[- a -(\frac{i}{2} +\frac {i\a}{4}) c-(i+\frac{i\a}{4})d
+(1+\frac{9\a}{4} +\frac{\a^2}{4})e \right. \right. \nonumber \\
&&~~~~~~~~~~~~~~ \left. \left.
-(\frac{\a}{4} + \frac{\a^2}{8})g
-\frac{i\a}{4} h -(\frac{\a}{2}+\frac{\a^2}{8})k \right]
\pa \Phi_1 \pa^2 \Phi_1  \right. \nonumber\\
&&~~~~ \left. + \left[ ia +(\frac{1}{2}-\frac{\a}{4})c -\frac{\a}{4}d
+(2i+\frac{3i\a}{4} -\frac{i\a^2}{4})e - (\frac{5i\a}{4}-\frac{i\a^2}{8})g
\right. \right. \nonumber \\
&&~~~~~~~~~~~~~~ \left. \left.
+(\frac{1}{2}- \frac{\a}{4})h -(i+i\a-\frac{i\a^2}{8})k \right] \pa
\Phi_2 \pa^2 \Phi_1 \right. \nonumber\\
&&~~~~ \left.+ \left[ ib-(1+\frac{\a}{4})c-(1+\frac{\a}{4})d +2ie
+(\frac{3i\a}{4} +\frac{i\a^2}{4})f \right.\right.
\nonumber \\
&&~~~~~~~~~~~~~~ \left. \left.
-(i+ \a +\frac{i\a^2}{8})g  + \frac{1}{2}h -(2i+\frac{5i\a}
{4}+\frac{i\a^2}{8})k
 \right] \pa \Phi_1 \pa^2 \Phi_2 \right. \nonumber\\
&&~~~~ \left. + \left[b+\frac{i\a}{4}c +\frac{i\a}{4}d +(1-\frac{9\a}{4}
+\frac{\a^2}{4})f+ (\frac{\a}{2} - \frac{\a^2}{8})g +
( \frac{\a}{4}- \frac{\a^2}{8})k \right] \pa\Phi_2 \pa^2 \Phi_2 \right.
\nonumber\\
&&~~~~ \left. +\left[ -\frac{i}{2}c -\frac{3\a}{4}f+(1+\frac{\a}{2})g
+(\frac{i}{2}-\frac{i\a}{8})h+(1+\frac{\a}{4})k \right]
D \Phi_1 D \pa^2 \Phi_2
\right. \nonumber \\
&&~~~~ \left.
+\left[\frac{i}{2}c -(3+\frac{3\a}{4})e + \frac{\a}{4}g + \frac{i\a}
{8}h +(1+\frac{\a}{2})k \right] D \Phi_2 D \pa^2 \Phi_1 \right. \nonumber \\
&&~~~~ \left.
+ \left[ id+3e+ \frac{\a}{2}g + (\frac{i}{2}-\frac{i\a}{2})h -
(1+\frac{\a}{2})k \right] D \pa \Phi_1 D \pa \Phi_2  \right. \nonumber \\
&&~~~~ \left.
+ \left[ -3ie+ig-\frac{1}{2}h +2ik \right] \pa \Phi_2 D \pa \Phi_1 D \Phi_2
+ \left[2e+2f-2g-2k \right] (\pa \Phi_2)^2 \pa \Phi_1 \right. \nonumber \\
&&~~~~ \left.
+\left[ -3if+2ig+\frac{1}{2}h+ik \right] \pa \Phi_2 D \pa \Phi_2 D \Phi_1
\right) e^{\Phi_1-i\Phi_2} \nonumber \\
&&
\ena
The third exponential needs not be considered because the theory is symmetric
under $\Phi_1 \rightarrow - \Phi_1$.

Up to D-derivatives the terms in the r.h.s. of eqs.(22) and (23)
are all independent and
they are not total derivatives. Therefore
the $J^{(3)}$ current will satisfy the conservation equation (19) if the
various coefficients separately vanish.
This leads to a set of equations for
$a,b,\dots ,k$ which can be solved {\em nontrivially}. Thus we
obtain, up to an overall normalization factor, the {\cal quantum} spin-$3$
conserved current
\bea
J^{(3)} &=& i \left(-4 + \frac{13}{8\pi} \b^2 + \frac{7}{32\pi^2}
{\b}^4 - \frac{{\b}^6}{64\pi^3}  \right) D \partial \Phi_1 \partial^2
\Phi_1  \nonumber \\
&-& i \left(1 - \frac{7}{8\pi} \b^2 - \frac{7}{32\pi^2} \b^4 +
\frac{\b^6}{64\pi^3} \right) D \partial \Phi_2 \partial^2 \Phi_2  \nonumber \\
&+& \left(-6 + \frac{3}{2\pi}\b^2 \right)D \Phi_1 \partial \Phi_1 \partial^2
\Phi_2 + \left(6 - \frac{9}{4\pi} \b^2 + \frac{3}{16\pi^2}\b^4 \right)
(\partial \Phi_1)^2 D \partial \Phi_2  \nonumber \\
&-& i \left(1 - \frac{\b^2}{2\pi}
\right) (\partial \Phi_1)^3 D \Phi_1 +
i \left(1 + \frac{\b^2}{4\pi} \right) (\partial \Phi_2)^3 D \Phi_2
\nonumber \\
&-& 6 D \Phi_1 D \partial \Phi_1 D \Phi_2 \partial \Phi_2  +
\frac{3i}{4\pi}\b^2
(\partial \Phi_1)^2 \partial \Phi_2 D \Phi_2
\ena
The classical current is obtained by setting
$\b^2 = 0$. It is worth noticing that at the classical level
$J^{(3)}$ is equivalent,
up to total $D$-derivative terms, to the classical current
$W^{(1)}DW^{(1)}+2 W^{(1)}
W^{(\frac{3}{2})}$. While in the conformal theory the two terms are separately
holomorphic currents, in the affine case only the linear combination above
satisfies the classical conservation law. At the quantum level we could not
use the renormalized expressions in eqs.(14) and (15) since $J^{(3)}$ is a
composite operator and the explicit calculation in eqs.(22), (23) was necessary
in order to determine its complete renormalized form.

The corresponding spin-$3$ charge
\EQ
Q^{(3)} = \int dz d \theta J^{(3)}(z,\theta)
\EN
satisfies $\bar{D} Q^{(3)}=0$ and it commutes with the Hamiltonian
of the system. Single particle states
are eigenstates of the charge operator with eigenvalues proportional to the
particle charge $\omega$, defined as $Q^{(3)}|p_j \rangle=
\b^2 \o_j p^3_{+j} |p_j \rangle$.
(We follow here the notations of Ref. \cite{dgz}.)
For an on-shell 3-point correlation function
$\langle \Phi_a \Phi_b \Phi_c \rangle$  momentum and charge conservations
lead to
\EQ
p_{+a} + p_{+b} + p_{+c} = 0~~~~~~~~~,~~~~~~~~~
\omega_a p_{+a}^3 + \omega_b p_{+b}^3 + \omega_c p_{+c}^3 = 0
\EN
Specializing these relations to the vertex function
$\langle \Phi_1 \Phi_1 \Phi_2 \rangle$  of the $B^{(1)}(1,1)$ theory and
writing momenta in terms of rapidities ($p_{+j} = \frac{m_j}{\sqrt{2}}
e^{\Theta_j}$),
in a frame of reference where one of the particle has rapidity zero and
the other two $\pm i \Theta$, we obtain
\EQ
2 m_1 \cos{\Theta} = m_2~~~~~~,~~~~~~
2 \omega_1 m_1^3 \cos{3 \Theta} = \omega_2 m_2^3
\EN
Therefore the charges and masses of the theory must satisfy
\EQ
\frac{\omega_2}{\omega_1} = 1 - 3 \frac{m_1^2}{m_2^2}
\label{8}
\EN
We verify up to one-loop level that this relation is indeed valid.

The charges $\omega_j$, $j=1,2$ can be computed in terms of the
on-shell matrix elements $\langle p_j|J^{(3)}(0)|p_j \rangle$
(see Ref. \cite{dgz} for details on the general procedure).
At the classical level they are obtained from
the quadratic terms in $J^{(3)}$. One finds $\omega^{(0)}_1 = 4$,
$\omega^{(0)}_2 = 1$ so that using the classical mass ratio
$m_1^2/m_2^2 =1/4$, eq.(28) is satisfied.
We compute now first order
corrections to the masses and the charges.
One-loop corrections to the mass spectrum are given by on-shell
self-energy supergraphs with massive propagators
\EQ
\langle \Phi_i(Z,\bar{Z}) \Phi_j(0,0) \rangle =- i \delta_{ij} \b^2
\frac{( \bar{D}D + M_i)}{\Box + 2 M_i^2} \delta^{(2)}(\theta)
\EN
We obtain the following contributions to the effective action
($e^{iS} \rightarrow e^{i\G}$)
\bea
&(A):& \qquad -2 \int d^2 \theta~ \Phi_1
( \bar{D}D + M_1 + M_2) \Phi_1 ~\S(p^2;m^2_1,m^2_2)
\\ \nonumber
&(B):& \qquad -\int d^2 \theta~ \Phi_2 ( \bar{D}D +2M_1) \Phi_2 ~
\S(p^2;m_1^2,m_1^2)
\ena
where
\EQ
\S(p^2;m^2_i,m^2_j) = \frac{1}{(2{\pi})^2} \int \frac{d^2 k }
{(k^2-m^2_i)[(k-p)^2-m^2_j]}
\EN
We note that when evaluated on-shell with $p^2_2=8=(2m_1)^2$,
$\S(p^2_2;m^2_1,m^2_1)$ is divergent since the two particles
are at threshold. In
any event using the on-shell conditions $
 \bar{D}D \Phi_j = M_j \Phi_j$
the kinematic factors in $(A)$ and $(B)$ vanish so that
at one-loop the classical masses are not renormalized.

In order to compute one-loop corrections to the
$\omega$-charges we need evaluate
the one-loop supergraphs with one insertion of the current $J^{(3)}$ as
shown in Fig.$2$.
The relevant terms in the current are
\bea
J^{(3)} \sim && i(-4+\frac{13}{8\pi}\b^2) D\pa \Phi_1 \pa^2 \Phi_1 +i
(-1+\frac{7}{8\pi}\b^2) D\pa \Phi_2 \pa^2 \Phi_2 \nonumber \\
&& -6D\Phi_1 \pa \Phi_1 \pa^2 \Phi_2 +6(\pa\Phi_1)^2 D\pa \Phi_2
\ena
The diagrams in Fig.$2a,b$ correspond to one-loop contributions from the
classical part of the current, while Fig.$2c$ corresponds
to tree-level contributions from the $O(\b^2)$
terms in the current and to wave-function renormalization corrections.
As usual, in order to compute the supergraphs one first perform the D-algebra
in the loop and then evaluates the momentum integrals.
We obtain, for $\o_1$
\EQ
(a)~~:~~ -\frac{3}{4\pi} \b^2~~~~~~,~~~~~~(b)~~:~~-\frac{1}{8\pi}\b^2~~~~~~~,
{}~~~~~~(c)~~:~~ (-\frac{13}{4} +1)\frac{\b^2}{2\pi}
\EN
where in (c) the two terms correspond to the $O(\b^2)$ current insertion and to
the contribution from wave-function renormalization, respectively.
For $\o_2$, the one-loop corrections are
\EQ
(a)~~:~~\frac{3}{4\pi}\b^2~~~~~~,~~~~~~(b)~~:~~(-\frac{3}{4} -\frac{1}{16}I_0)
\frac{\b^2}{2\pi}
{}~~~~~~(c)~~:~~(-\frac{7}{4} +\frac{1}{16} I_0)\frac{\b^2}{2\pi}
\EN
where again in (c) the two types of contributions are listed separately
and $I_0$ denotes the divergent integral
\EQ
I_0 = \int_0^1  \frac{dx}{(x-\frac{1}{2})^2}
\EN
which arises from threshold effects in the computation of $\S(8;2,2)$.
Summing the terms in $(a)$, $(b)$ and $(c)$ the divergence cancels and we find
that the $O(\b^2)$ corrections maintain the classical
charge ratio $\o_2/
\o_1 = 1/4$, in agreement with the absence of mass renormalization.

As a further consistency check, we have computed the on-shell vertex function
$\langle \Phi_2 \Phi_2 \Phi_2 \rangle$ at one-loop and obtained
a zero result (the corresponding $\langle \Phi_1 \Phi_1 \Phi_1 \rangle$
corrections are absent because of the coupling). Indeed the vanishing of
the $\Phi^3_1$ and $\Phi^3_2$ couplings is required by eq.(26) (with $a=b=c$)
and the fact that the charges $\o_1$ and $\o_2$
have a non-vanishing one-loop correction.

We conclude summarizing our results: using massless perturbation in $N=1$
two-dimensional superspace we have constructed the quantum W-supercurrents
of the supersymmetric Toda theory based on the Lie superalgebra $B(1,1)$.
The system possesses two independent holomorphic currents, the stress-energy
tensor $W^{(1)}$ and $W^{(\frac{3}{2})}$ whose first component is fermionic.
Then we have considered the supersymmetric affine extension of the model.
The addition of the perturbation is such that the first nontrivial conserved
current appears now at $s=3$. We have obtained its renormalized expression
 to all-loop order and computed the corresponding charge up to one-loop.
The $B^{(1)}(1,1)$ theory is not manifestly unitary, nonetheless its particle
mass spectrum is real and not renormalized to lowest order in perturbation
theory.

Finally we observe that the action in eq.(16)
reduces to the supersymmetric
sine-Gordon model if we set $\Phi_1=0$. Therefore the spectrum of the theory
contains
soliton solutions which are given by $\Phi_1=0$ and $\Phi_2$ assuming the
field configurations of the super sine-Gordon solitons \cite{sol}. Here
again the masses are {\em real}. Since the system is integrable one might hope
to be able to construct the corresponding S-matrix and determine if and when
the non-unitary sector decouples.

\vskip 6pt

Further details and the extension to the other supersymmetric Toda theories
will be presented in a separate publication \cite{gpz2}.

\newpage


\begin{center}
\begin{tabular}{lclc}
$B(1,1)$:
\setlength{\unitlength}{1.5pt}
\begin{picture}(120,90)
\thicklines
\put(15,5){\circle{12}}
\put(11.5,1.5){\line(1,1){7}}
\put(18.5,1.5){\line(-1,1){7}}
\put(20,8){\line(1,0){26}}
\put(20,2){\line(1,0){26}}
\put(36,5){\line(-1,1){7}}
\put(29,-2){\line(1,1){7}}
\put(50,5){\circle*{15}}
\end{picture}

$B^{(1)}(1,1)$:
\setlength{\unitlength}{1.5pt}
\begin{picture}(120,90)
\thicklines
\put(15,-12){\circle{12}}
\put(11.5,-15.5){\line(1,1){7}}
\put(18.5,-15.5){\line(-1,1){7}}
\put(12.5,-7){\line(0,1){24}}
\put(17,-7){\line(0,1){24}}
\put(15,22){\circle{12}}
\put(11.5,18.5){\line(1,1){7}}
\put(18.5,18.5){\line(-1,1){7}}
\put(18,-7){\line(1,1){11}}
\put(20,-11){\line(1,1){12}}
\put(32,5){\circle*{15}}
\put(30,10){\line(-1,1){10}}
\put(27,7){\line(-1,1){10}}
\put(25,12){\line(0,1){7}}
\put(18,12){\line(1,0){7}}
\put(25,-10){\line(0,1){7}}
\put(18,-3){\line(1,0){7}}
\put(7,22){\makebox(0,0){$1$}}
\put(7,-12){\makebox(0,0){$1$}}
\put(40,5){\makebox(0,0){$2$}}
\end{picture}

\end{tabular}
\end{center}
\vspace{1.5cm}
\begin{center}
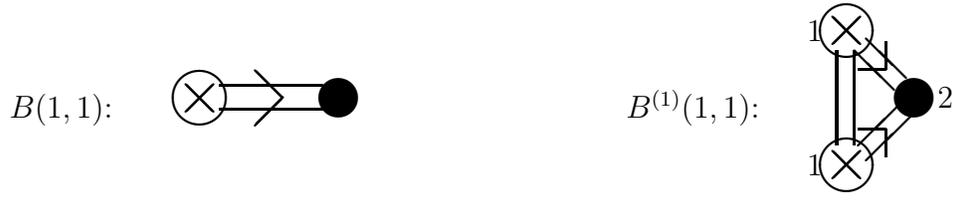
{\bf Figure 1:} Fermionic Dynkin diagrams
\end{center}


\begin{center}
\begin{tabular}{lclc}

\setlength{\unitlength}{1.5pt}
\begin{picture}(120,90)
\thicklines
\put(13,5){\oval(5,5)[tr]}
\put(13,6){\oval(5,5)[br]}
\put(14,1){\oval(5,5)[tl]}
\put(14,2){\oval(5,5)[bl]}
\put(13,-3){\oval(5,5)[tr]}
\put(13,-2){\oval(5,5)[br]}
\put(14,-7){\oval(5,5)[tl]}
\put(14,-6){\oval(5,5)[bl]}
\put(13,-11){\oval(5,5)[tr]}
\put(13,-10){\oval(5,5)[br]}
\put(14,-15){\oval(5,5)[tl]}
\put(14,-14){\oval(5,5)[bl]}
\put(13,-19){\oval(5,5)[tr]}
\put(28,-19){\circle{26}}
\put(40,-19){\line(1,0){25}}
\put(-9,-19){\line(1,0){25}}
\put(28,-47){\makebox(0,0){a)}}
\end{picture}

\setlength{\unitlength}{1.5pt}
\begin{picture}(120,90)
\thicklines
\put(28,11){\oval(5,5)[tr]}
\put(28,12){\oval(5,5)[br]}
\put(29,7){\oval(5,5)[tl]}
\put(29,8){\oval(5,5)[bl]}
\put(28,3){\oval(5,5)[tr]}
\put(28,4){\oval(5,5)[br]}
\put(29,-1){\oval(5,5)[tl]}
\put(29,0){\oval(5,5)[bl]}
\put(28,-5){\oval(5,5)[tr]}
\put(28,-4){\oval(5,5)[br]}
\put(28,-19){\circle{26}}
\put(40,-19){\line(1,0){25}}
\put(-9,-19){\line(1,0){25}}
\put(28,-47){\makebox(0,0){b)}}
\end{picture}

\setlength{\unitlength}{1.5pt}
\begin{picture}(120,90)
\thicklines
\put(28,3){\oval(5,5)[tr]}
\put(28,4){\oval(5,5)[br]}
\put(29,-1){\oval(5,5)[tl]}
\put(29,0){\oval(5,5)[bl]}
\put(28,-5){\oval(5,5)[tr]}
\put(28,-4){\oval(5,5)[br]}
\put(29,-9){\oval(5,5)[tl]}
\put(29,-8){\oval(5,5)[bl]}
\put(28,-13){\oval(5,5)[tr]}
\put(28,-12){\oval(5,5)[br]}
\put(29,-17){\oval(5,5)[tl]}
\put(29,-16){\oval(5,5)[bl]}
\put(4,-19){\line(1,0){50}}
\put(26,-22){\line(1,1){7}}
\put(32,-22){\line(-1,1){7}}
\put(28,-47){\makebox(0,0){c)}}
\end{picture}

\end{tabular}
\end{center}
\vspace{1.3in}
\begin{center}
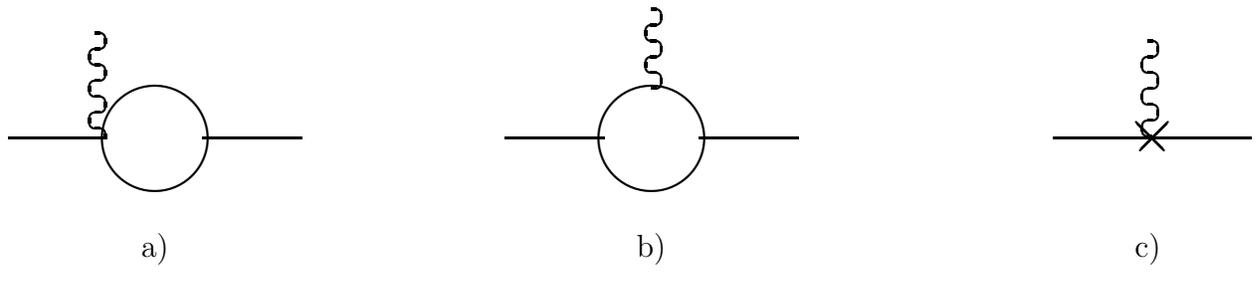
{\bf Figure 2:} Diagrams for the calculation of the charge; the wavy line
indicates the
\end{center}
{}~~~~~~~~~~~~~~~~~~insertion of the current.

\end{document}